\documentstyle[preprint,pra,aps]{revtex}

\def\sqr#1#2{{\vcenter{\hrule height.#2pt\hbox{\vrule width.#2pt
height#1pt \kern#1pt \vrule width.#2pt}\hrule height.#2pt}}}

\newcommand{\be}{\begin{equation}}
\newcommand{\ee}{\end{equation}}
\newcommand{\ba}{\begin{eqnarray}}
\newcommand{\ea}{\end{eqnarray}}

\draft

\begin{document}
\draft
\title{Equipartition and Virial theorems in a nonextensive optimal Lagrange multipliers
scenario}
\author{ S. Mart\'{\i}nez${^{1,\,2}}$\thanks{%
E-mail: martinez@venus.fisica.unlp.edu.ar}, F. Pennini${%
^{1,\,2}}$\thanks{
E-mail: pennini@venus.fisica.unlp.edu.ar}, and A. Plastino${^{1,\,2}}$%
\thanks{
E-mail: plastino@venus.fisica.unlp.edu.ar}}
\address{$^1$ Physics Department, National University La Plata, C.C. 727 ,\\
1900 La Plata, Argentina.}
\address{$^2$ Argentina National Research Council (CONICET)}

\maketitle

\begin{abstract}
We revisit some topics of classical thermostatistics from the
perspective of the nonextensive optimal Lagrange multipliers
(OLM), a recently introduced technique for dealing with the
maximization of Tsallis' information measure. It is shown that
Equipartition and Virial theorems can be reproduced by Tsallis'
nonextensive formalism independently of the value of the
nonextensivity index. \vskip 3mm PACS: 05.30.-d,05.70.Ce. \vskip
3mm

\vspace{1 cm}
\end{abstract}

\newpage 

\section{Introduction}

Tsallis' thermostatistics (TT) \cite{t01,t1,review,t03,t3} is by
now recognized as a new paradigm for statistical mechanical
considerations. It revolves around the concept of Tsallis'
information measure $S_q$, a generalization of Shannon's one, that
depends upon a real index $q$ and becomes  Shannon's measure
\cite{t01} for the particular value $q=1$.

  The Equipartition theorem, first formulated  by Boltzmann in 1871, and
  the Virial one, due to Clausius (1870) \cite{pathria} are two pillars of classical
physics. They were discussed within the TT framework in
\cite{PPT94}, in terms of the Curado-Tsallis nonextensive
normalizing treatment \cite{t3}, and also in \cite{PL99}, by
recourse to escort probabilities \cite{mendes}. In the present
work we are going to revisit these classical subjects in
connection with a recently advanced scheme for dealing with
Tsallis' thermostatistics \cite{olm,ley0}, that seems to yield
illuminating insights into classical themes.

Tsallis' thermostatistics involves extremization of Tsallis'
entropy by recourse to the celebrated technique of Lagrange. A key
TT ingredient is the particular way in which expectation values
are computed \cite{mendes} and, in such a respect, several
proposals have been considered during the last ten years
\cite{review}. No matter which recipe one chooses (for $q\ne 1$),
classical phenomena are always reproduced by TT in the limit
$q\rightarrow 1$ \cite{review}. Recently, a new algorithm (to be
henceforth referred to as the OLM-one) has been advanced to such
an extremizing end that diagonalizes the Hessian associated to the
Lagrange procedure \cite{olm} and yields the optimal Lagrange
multipliers associated to the input expectation values. A diagonal
Hessian enormously facilitates ascertaining just what kind of
extreme the Lagrange method is leading to \cite{olm}.

A key point in TT considerations is the following one: the entropy
constant $k$ is usually identified with Boltzmann's $k_{B}$.
However, the only
certified fact one can be sure of is ``$k\rightarrow k_{B}$ for $%
q\rightarrow 1$'' \cite{review}, which entails that there is room {\it to
choose $k=k(q)$ in any suitable way}. It is seen \cite{olm} that if one
chooses

\begin{equation}
k=k_{B}\bar{Z}_{q}^{q-1},  \label{k}
\end{equation}
where $\bar{Z}_{q}$ stands for the partition function,
in conjunction with the OLM formalism \cite{olm}, the
classical harmonic oscillator determines a specific
heat $C_{q}=k_{B}$, so that {\it the classical Gibbs'
result arises without the need to invoke the limit}
$q\rightarrow 1$. Additionally, this treatment is able
i) to reproduce thermodynamics' zero-th law
\cite{ley0}, a feat that had eluded previous TT
practitioners, and ii) the classical $q$-independent
value for the mean energy of the ideal gas \cite{gas}.
It is then not unreasonable to conjecture that many
other classical results might be obtained by TT without
invoking the $q\rightarrow 1$ limit.

In the present effort we re-discuss the equipartition theorem and
the virial one from an OLM viewpoint. We also revisit the ideal
gas problem as an application. More precisely, we perform an OLM
treatment of the subject following the canonical ensemble
strictures.

\section{Brief review of OLM formalism in a classical scenario}
 The most general classical treatment
requires consideration of the probability density $p({\bf x})$
that maximizes Tsallis' entropy \cite{t01,t1,t2}

\begin{equation}
\frac{S_{q}}{k}=\frac{1-\int d{\bf x\ }p^{q}({\bf x})}{q-1},
\label{entropia}
\end{equation}
by recourse to the Lagrange's technique, subject to the
foreknowledge of $M$ generalized expectation values \cite{mendes}
\begin{equation}
\left\langle \left\langle O_{j}\right\rangle \right\rangle _{q}=\frac{\int d%
{\bf x\ }p^{q}({\bf x})O_{j}({\bf x})}{\int d{\bf x\ }p^{q}({\bf
x})}, \label{gener}
\end{equation}
where $O_{j}({\bf x})$ ($j=1,\ldots ,M$) denote the $M$ relevant
dynamical quantities (the observation level \cite{aleman}), $q\in
\Re $ is Tsallis' nonextensivity index, ${\bf x}$ is a phase space
element ($N$ particles in a $D$-dimensional space), and $k$ is the
entropy constant, akin to the famous Boltzmann one $k_B$, employed
in the orthodox statistics.

Tsallis' normalized probability distribution
\cite{mendes}, is obtained by following the well known
MaxEnt route \cite{katz}. Instead of effecting the
variational treatment of \cite{mendes}, involving $M+1$
Lagrange multipliers $\lambda_j$ (associated to
 constraints given by the normalization condition
together with the $M$ equations (\ref{gener})), the OLM
technique follows an alternative path \cite{olm} with
Lagrange multipliers $\lambda^{\prime}_j$: one
maximizes Tsallis' generalized entropy $S_q$
(\ref{entropia}) \cite{t01,t1,t2} subject to the {\it
modified} constraints (``centered" generalized
expectation values) \cite{t01,olm}

\begin{eqnarray}
\int d{\bf x\ }p({\bf x})-1 &=&0 \\ \int d{\bf x}\ p({\bf
x})^{q}\left( O_{j}({\bf x})-\left\langle \left\langle
O_{j}\right\rangle \right\rangle _{q}\right) &=&0.
\label{vinculos}
\end{eqnarray}

 The resulting
probability distribution reads \cite{olm}

\begin{equation}
p({\bf x})=\frac{f({\bf x})^{\frac{1}{1-q}}}{\bar{Z}_{q}},  \label{px}
\end{equation}
where

\begin{equation}
\bar{Z}_{q}=\int d{\bf x\ }f({\bf x})^{\frac{1}{1-q}}.  \label{Zqp}
\end{equation}
and
\begin{equation}
f({\bf x})=1-(1-q)\sum_{j}^{M}\,\lambda _{j}^{\prime }\left( O_{j}({\bf x}%
)-\left\langle \left\langle O_{j}\right\rangle \right\rangle _{q}\right)
\label{carac}
\end{equation}
is the so-called configurational characteristic.

Although the Tsallis-Mendes-Plastino (TMP) procedure originally
devised in \cite{mendes} overcomes most of the problems posed by
the old, unnormalized way of evaluating Tsallis' generalized mean
values \cite{mendes,pennini}, it yields probability distributions
that are self-referential, which entails some numerical
difficulties. The complementary OLM treatment of \cite{olm}
surmounts these hardships. Inspection of (\ref {px}) shows that
the self-reference problem has vanished.

One shows in \cite{olm} that the relation
\begin{equation}
\int d{\bf x\ }p^{q}({\bf x})=\bar{Z}_{q}^{1-q},  \label{relac1}
\end{equation}
valid under TMP, still holds. Eq. (\ref{relac1}) allows one to
connect the Lagrange multipliers $\lambda _j$ of the TMP procedure
\cite{mendes} with the corresponding OLM $\lambda _j^{\prime }$
via \cite{olm}
\begin{equation}  \label{lambda'}
\lambda_j^{\prime }= \frac{\lambda_j}{\bar{Z}_{q}^{1-q}},
\end{equation}
and to write the entropy as \cite{olm}
\begin{equation}
S_{q}=k\;{\rm \ln }_{q}\bar{Z}_{q},  \label{S2}
\end{equation}
where ${\rm \ln }_{q}x=(1-x_{q}^{1-q})/(q-1)$ has been used.

 If now, following \cite{mendes} we define
\begin{equation}
\ln _{q}Z_{q}^{\prime }={\rm \ln }_{q}\bar{Z}_{q}-\sum_{j}\lambda
_{j}^{\prime }\ \left\langle \left\langle O_{j}\right\rangle \right\rangle
_{q},  \label{lnqz'}
\end{equation}
and additionally use \cite{olm}
\begin{equation}
k^{\prime }=k\;\bar{Z}_{q}^{1-q},  \label{k'}
\end{equation}
then \cite{olm}
\begin{eqnarray}
\frac{\partial S_{q}}{\partial \left\langle \left\langle O_{j}\right\rangle
\right\rangle _{q}} &=&k^{\prime }\lambda _{j}^{\prime }  \label{termo1} \\
\frac{\partial }{\partial \lambda _{j}^{\prime }}\left( \ln
_{q}Z_{q}^{\prime }\right) &=&-\left\langle \left\langle O_{j}\right\rangle
\right\rangle _{q}.  \label{termo2}
\end{eqnarray}

Equations (\ref{termo1}) and (\ref{termo2}) constitute the basic
Information Theory relations to build up Statistical Mechanics
{\em \`a la} Jaynes \cite{katz}. Notice that, when $k$ is given by
(\ref{k}), Equation (\ref{k'}) leads to $k^{\prime}=k_B$
\cite{t01}.

 Remembering that \cite{mendes}
\be
\frac{\partial S_{q}}{\partial \left\langle \left\langle
O_{j}\right\rangle \right\rangle _{q}} =k\lambda _{j},
\ee
one is
straightforwardly led to \cite{olm}

\begin{equation}
k^{\prime }\lambda _j^{\prime }=k\lambda _j,  \label{rel}
\end{equation}
which entails that the intensive variables are identical in both
alternative pictures, TMP and OLM \cite{olm}.

As a special instance of Eq. (\ref{rel}), for the Canonical
Ensemble it adopts the appearance
\be
k^{\prime }\beta ^{\prime } =k\beta =\frac %
1T.  \label{C1} \ee

Looking at Eq. (\ref{C1}) one gathers that the temperature $T$ is
the same for both sets of Lagrange multipliers.

\label{NTT}

Before tackling the equipartition theorem, some preliminary results are
needed, that we discuss next.

\section{Normalization considerations}

Replacing (\ref{px}) into (\ref{relac1}) we obtain for the partition
function the expression
\begin{equation}
\bar{Z}_{q}=\int d{\bf x\ }f({\bf x})^{\frac{q}{1-q}},  \label{Zqp2}
\end{equation}
and, comparing it to (\ref{Zqp}), we deduce the (at first sight surprising)
relationship
\begin{equation}
{\cal D} = \int d{\bf x\ }f({\bf x})^{\frac{q}{1-q}}-\int d{\bf x\ }f({\bf x}%
)^{\frac{1}{1-q}}=0,  \label{rel1}
\end{equation}
valid for all $q$. That this is indeed so can also be gathered by first
recasting (\ref{rel1}) in the form
\begin{equation}
{\cal D} = \int d{\bf x}\left[ f({\bf x})^{\frac{q}{1-q}}\left( 1-f({\bf x}%
)\right) \right],
\end{equation}
and then using Eq. (\ref{carac}) to obtain
\begin{equation}
{\cal D}= (1-q)\sum_{j}\lambda _{j}^{\prime } \int d{\bf x}f^{\frac{q}{1-q}%
}\left( \widehat{O}_{j}-\left\langle \left\langle \widehat{O}%
_{j}\right\rangle \right\rangle _{q}\right),
\end{equation}
that, remembering (\ref{px}), has to vanish identically on account of (\ref
{vinculos}).

We introduce now the factor
\begin{equation}
F(q)\equiv \frac{\int d{\bf x}f({\bf x})^{\frac{1}{1-q}}}{\int d{\bf x}f(%
{\bf x})^{\frac{q}{1-q}}}=1,  \label{factor}
\end{equation}
in order to be in a position to use it later on.

It is to be stressed that $F(q)=1$ only within the so-called ``normalized"
framework of \cite{mendes}. If one uses, instead, the Curado-Tsallis
normalization \cite{t3}, this is not so ($F(q) \ne 1$). The OLM formalism
does employ the normalized treatment. It must be pointed out that the $F(q)=1
$-result has also been obtained within a purely quantal, Green function
scheme by Lenzi, Mendes, and Rajagopal \cite{lenzi}.

\section{Generalized equipartition and virial theorems}

In classical statistical physics, a Hamiltonian dynamical system is
described by an appropriate phase space probability distribution $%
p(r_{i},p_{i}).$ Just one assumption will be made on
the probability density (PD): that it depend on the
phase space variables $(r_{i},p_{i})$ only through the
Hamiltonian $H(r_{i},p_{i})$.

Tsallis' normalized probability distribution \cite{mendes} is
obtained by maximizing Tsallis' generalized entropy $S_{q}$ given
by Eq. (\ref{entropia}), subject to the constraints \cite{t01,olm}

\begin{eqnarray}
\int d\Omega p(r_{i},p_{i}) &=&1  \label{vinculos2} \\
\int d\Omega p^{q}(r_{i},p_{i})\left( H(r_{i},p_{i})-\left\langle
\left\langle H\right\rangle \right\rangle _{q}\right)  &=&0,
\end{eqnarray}
where the generalized expectation values \cite{mendes}

\begin{equation}
\left\langle \left\langle H\right\rangle \right\rangle _{q}=\frac{\int
d\Omega p^{q}(r_{i},p_{i})H(r_{i},p_{i})}{\int d\Omega p^{q}(r_{i},p_{i})}
\label{gener2}
\end{equation}
are assumed to be a priori known. They constitute the macroscopic
information at hand concerning the system. $d\Omega $ stands for the
corresponding phase space volume element

\begin{equation}
d\Omega =(1/(N!h^{DN}))\prod_{i=1}^{DN}dr_{i}dp_{i},\,\,\,(i=1\ldots DN),
\end{equation}
where $h$ is the linear dimension (i.e. the size) of the
elementary cell in phase space, and we assume $\int
\prod_{i=1}^{DN}dr_{i}=V^{N}$ with $V$ the system's  volume.

The resulting probability distribution reads \cite{olm}

\begin{equation}
p(r_{i},p_{i})=\bar{Z}_{q}^{-1}\left[ 1-(1-q)\beta ^{\prime }\left(
H(r_{i},p_{i})-\left\langle \left\langle H\right\rangle \right\rangle
_{q}\right) \right] ^{\frac{1}{1-q}},  \label{rho2}
\end{equation}
where

\begin{equation}
\bar{Z}_{q}=\int d\Omega \left[ 1-(1-q)\beta ^{\prime }\left(
H(r_{i},p_{i})-\left\langle \left\langle H\right\rangle \right\rangle
_{q}\right) \right] ^{\frac{1}{1-q}}.  \label{Zqpc}
\end{equation}

Let $A(q_{i,}p_{i})$ denote a generic dynamical quantity. Its generalized
mean value is given by
\begin{equation}
\left\langle \langle A\rangle \right\rangle _{q}=\frac{\int d\Omega \
A(r_{i,}p_{i})\left[ 1-\beta ^{^{\prime }}(1-q)\left( H-\left\langle
\left\langle H\right\rangle \right\rangle _{q}\right) \right] ^{\frac{q}{1-q}%
}}{\int d\Omega \left[ 1-\beta ^{^{\prime }}(1-q)\left( H-\left\langle
\left\langle H\right\rangle \right\rangle _{q}\right) \right] ^{\frac{q}{1-q}%
}}.
\end{equation}

Let us now assume that our Hamiltonian $H$ can be split into two
pieces according to
\begin{equation}
H=g_0+g  \label{H}
\end{equation}
where $g$ is a homogeneous function of degree $\gamma $ of $L$ canonical
variables of which, say $\nu $, are generalized coordinates while the
remaining ones are $\mu =L-\nu $ generalized momenta
\begin{equation}
g=g(r_{_{1}},\ldots ,r_{\nu },p_{1},\ldots ,p_{\mu }),  \label{g}
\end{equation}
and $g_0$ does not depend upon these variables. According to
Euler's theorem \cite{Fleming} we have
\be
\gamma g=\sum_{i=1}^{\nu }r_{i}(\partial g/\partial
r_{i})+\sum_{i=1}^{\mu }p_{i}(\partial g/\partial p_{i}) \ee so
that the generalized mean value of $g$ reads
\begin{equation}
\left\langle \left\langle g\right\rangle \right\rangle _{q}=\frac{1}{\gamma }%
\left[ \sum_{i=1}^{\nu }\langle \langle r_{i}(\partial g/\partial
r_{i})\rangle\rangle _{q}+\sum_{i=1}^{\mu } \langle \langle
p_{i}(\partial g/\partial p_{i})\rangle\rangle _{q}\right] .
\label{gq}
\end{equation}

We shall now discuss in some detail one generic term of this equation.
Consider
\begin{equation}
\left\langle \left\langle r_{k}(\partial g/\partial
r_{k})\right\rangle \right\rangle _{q}=\frac{\int d\Omega \
r_{k}(\partial g/\partial r_{k})\left[ 1-\beta ^{^{\prime
}}(1-q)\left( H-\left\langle \langle H\rangle \right\rangle
_{q}\right) \right] ^{\frac{q}{1-q}}}{\int d\Omega \left[ 1-\beta
^{^{\prime }}(1-q)\left( H-\left\langle \langle H\rangle
\right\rangle _{q}\right) \right] ^{\frac{q}{1-q}}},  \label{rk}
\end{equation}
 a multi-dimensional $(2DN)$ integral. Let us evaluate
the integral over $r_{k}$ ranging between $r_{a}$ and
$r_{b}$. These values are given by the well-known
Tsallis' cut-off condition \cite{review}: the
probability distribution vanishes in those regions of
phase space that would make (\ref{rho2}) a negative
quantity. For the numerator of (\ref{rk}) we have
\begin{equation}
J\equiv \int_{r_{a}}^{r_{b}}dr_{k}\ r_{k}(\partial g/\partial r_{k})\left[
1-\beta ^{^{\prime }}(1-q)\left( H-\left\langle \langle H\rangle
\right\rangle _{q}\right) \right] ^{\frac{q}{1-q}},  \label{jota}
\end{equation}
so that
\begin{equation}
\left\langle \left\langle r_{k}(\partial g/\partial r_{k})\right\rangle
\right\rangle _{q}=\frac{\int \ldots \int J\ dr_{1}\ldots
dr_{k-1}dr_{k+1}\ldots dp_{DN}}{\int d\Omega \left[ 1-\beta ^{^{\prime
}}(1-q)\left( H-\left\langle \langle H\rangle \right\rangle _{q}\right)
\right] ^{\frac{q}{1-q}}}.  \label{vmq}
\end{equation}

In order to obtain $J$ we proceed to an integration by parts. To
do so we first notice that, on account of Equations (\ref{H}) and
(\ref{g})
\begin{equation}
\frac{\partial }{\partial r_{k}}\left\{ \left[ 1-\beta ^{^{\prime
}}(1-q)\left( H-\left\langle \langle H\rangle \right\rangle _{q}\right)
\right] ^{\frac{1}{1-q}}\right\} =-\beta ^{^{\prime }}\frac{\partial g}{%
\partial r_{k}}\left[ 1-\beta ^{^{\prime }}(1-q)\left( H-\left\langle
\langle H\rangle \right\rangle _{q}\right) \right] ^{\frac{q}{1-q}},
\end{equation}
which, since the integrated part will vanish because of the above mentioned
cut-off condition, leads to
\begin{equation}
J=\frac{1}{\beta ^{^{\prime }}}\int dr_{k}\ \left[ 1-\beta ^{^{\prime
}}(1-q)\left( H-\left\langle \langle H\rangle \right\rangle _{q}\right)
\right] ^{\frac{1}{1-q}}.  \label{jota2}
\end{equation}

Insertion of (\ref{jota2}) into (\ref{vmq}) yields

\begin{equation}
\left\langle \left\langle r_{k}(\partial g/\partial r_{k})\right\rangle
\right\rangle _{q}=\frac{1}{\beta ^{^{\prime }}}\frac{\int d\Omega \left[
1-\beta ^{^{\prime }}(1-q)\left( H-\left\langle \langle H\rangle
\right\rangle _{q}\right) \right] ^{\frac{1}{1-q}}}{\int d\Omega \left[
1-\beta ^{^{\prime }}(1-q)\left( H-\left\langle \langle H\rangle
\right\rangle _{q}\right) \right] ^{\frac{q}{1-q}}},
\end{equation}
but, on account of (\ref{factor}) one has
\begin{equation}
\left\langle \left\langle r_{k}(\partial g/\partial r_{k})\right\rangle
\right\rangle _{q}=\frac{1}{\beta ^{^{\prime }}}.  \label{rq}
\end{equation}

It is apparent that each term in the sums appearing in (\ref{gq}) will yield
a contribution of the type (\ref{rq}), so that
\begin{equation}
\left\langle \left\langle g\right\rangle \right\rangle _{q}=\frac{L}{\gamma
\beta ^{^{\prime }}}=\frac{L}{\gamma }k_{B}T  \label{gqf}
\end{equation}
which is a generalized version of the equipartition
theorem \cite{pathria}. We have thus arrived to a new
an interesting result. {\it Contrary to current belief,
we see that the classical result is attained,
independently of the $q$-value}.

According to the canonical equations of motion, $\partial H/\partial r_{i}=-%
\dot{p}_{i}.$ Hence (\ref{rq}) leads to the statement
\begin{equation}
\left\langle \left\langle \sum_{i=1}^{DN}r_{i}\dot{p}_{i}\right\rangle
\right\rangle _{q}=-DNk_{B}T  \label{virial}
\end{equation}
which is Clausius' virial theorem \cite{pathria}. Again, {\it no
dependence upon $q$ is to be detected}.

As a simple application, consider now the classical ideal gas. We
necessarily reproduce the classical results, by virtue of the above
considerations: we deal with $N$ particles confined within a $D-$dimensional
box of volume $V$. Thermodynamical equilibrium at temperature $T$ is
assumed. The Hamiltonian is
\[
H=\sum_{i=1}^{N}\frac{{\bf p}_{i}^{2}}{2m}+\sum_{i=1}^{N}U_{wall}({\bf r}%
_{i}),
\]
where $U_{wall}$ is a contribution due to the constraints (walls of the
container). The constraint forces come into existence when the gas particles
collide with the walls. It thus follows that
\begin{equation}
\sum_{i=1}^{N}{\bf r}_{i}\cdot \frac{\partial H}{\partial {\bf r}_{i}}%
=-\sum_{i=1}^{N}{\bf r}_{i}\cdot {\bf F}_{wall}^{(i)}  \label{wall}
\end{equation}
where ${\bf F}_{wall}^{(i)}$ stands for the force on the i-particle due to
the walls of the box.

Following now a well-known argument \cite{pathria}, it is possible
then to express the last member of the above equation in terms of
the volume $V$ and
pressure $P$%
\begin{eqnarray}
-\sum_{i=1}^{N}{\bf r}_{i}\cdot {\bf F}_{wall}^{(i)} &=&-P\int_{wall}{\bf r}%
\cdot (-d{\bf s})=P\int_{wall}{\bf r}\cdot d{\bf s} \\
&=&P\int ({\bf \nabla }\cdot {\bf r})d^{3}r=DPV,  \nonumber
\end{eqnarray}
where $d{\bf s}$ denotes a (vector) surface element of the box in Eq. (\ref
{wall}). We get
\begin{equation}
\sum_{i=1}^{N}{\bf r}_{i}\cdot \frac{\partial H}{\partial {\bf r}_{i}}=DPV
\end{equation}

Now, by taking the canonical average and using the virial theorem
(unmodified by the nonextensive scenario) (\ref{virial}), we obtain
\begin{equation}
PV=Nk_{B}T,  \label{estado}
\end{equation}
i.e., the equation of state for the perfect gas. No $q$-dependence
is detected.

For the ideal gas \cite{PPT94,PL99,gas,Abe}, the total energy $E$
is a homogeneous quadratic function $(\gamma =2)$ of $DN$ momenta,
which allows one to write, according to (\ref{gqf})
\begin{equation}
\left\langle \left\langle H\right\rangle \right\rangle _{q}=\frac{1}{2}%
DNk_{B}T.  \label{HH}
\end{equation}

\section{Conclusions}

Classical thermostatistics has been the subject of the  present effort.

We have tackled some key issues, namely,
\begin{itemize}
\item Virial theorem,
\item Equipartition theorem,
\item Equation of state of the ideal gas,
\end{itemize}
and shown that they are, contrary to present belief, reproduced by
Tsallis' thermostatistics {\it independently of the value adopted
by the index $q$}. The present work lends further credence to  the
hypothesis that most important classical statistical results might
be reproduced by Tsallis' thermostatistics for all $q$-values.

\acknowledgements The financial support of the National Research Council
(CONICET) of Argentina is gratefully acknowledged. F. Pennini acknowledges
financial support from UNLP, Argentina.


\begin{references}
\bibitem{t01}  C. Tsallis, {\it Braz. J. of Phys.} {\bf 29} (1999) 1, and
references therein. See also
http://www.sbf.if.usp.br\-/WWW\_pages/Journals/BJP/Vol129/Num1/index.htm

\bibitem{t1}  C. Tsallis, {\it Chaos, Solitons, and Fractals} {\bf 6} (1995)
539, and references therein; an updated bibliography can be found in
http://tsallis.cat.cbpf.br/biblio.htm

\bibitem{review}  C. Tsallis, Nonextensive statistical mechanics and
thermodynamics: Historical background and present status, in ``Nonextensive
Statistical Mechanics and its Applications'', eds. S. Abe and Y. Okamoto,
``Lecture Notes in Physics'' (Springer-Verlag, Berlin, 2000), in press.

\bibitem{t03}  C. Tsallis, {\it Physics World 10} (July 1997) 42.

\bibitem{t3}  E. M. F. Curado and C. Tsallis, {\it J. Phys. A} {\bf 24}, L69
(1991); Corrigenda: {\bf 24}, 3187 (1991) and {\bf 25}, 1019 (1992).

\bibitem{pathria}  R. K. Pathria, Statistical Mechanics, Pergamon, New York,
1985.

\bibitem{PPT94}  A. R. Plastino, A. Plastino and C. Tsallis, {\it J. Phys.
A: Math. Gen.} {\bf 27} (1994) 5707.

\bibitem{PL99}  A. R. Plastino, J. A. S. Lima, {\it Physics Letters A} {\bf 260%
}  (1999) 46.


\bibitem{mendes}  C. Tsallis, R. S. Mendes, and A. R. Plastino, {\it Physica
A} {\bf 261}, 534 (1998).



\bibitem{olm}  S. Mart\'{\i }nez, F. Nicol\'{a}s, F. Pennini and A.
Plastino, preprint (2000) [physics/0003098].

\bibitem{ley0}  S. Mart\'{\i }nez, F. Pennini and A. Plastino, preprint
(2000) [cond-math/0004448].

\bibitem{gas}  S. Abe, S. Mart\'{\i }nez, F. Pennini and A. Plastino,
preprint (2000) [cond-mat/0006109].


\bibitem{t2}  C. Tsallis, {\it J. Stat. Phys.} {\bf 52}, 479 (1988).

\bibitem{aleman}  E. Fick and G. Sauerman, {\it The quantum statistics of
dynamic processes} (Springer-Verlag, Berlin, 1990).

\bibitem{katz}  E. T. Jaynes in {\it Statistical Physics}, ed. W. K. Ford
(Benjamin, NY, 1963); A. Katz, {\it Statistical Mechanics},
(Freeman, San Francisco, 1967).

\bibitem{pennini}  F. Pennini, A. R. Plastino and A. Plastino, {\it Physica
A } {\bf 258}, 446 (1998).


\bibitem{lenzi}  E. K. Lenzi, R. S. Mendes, and A. K. Rajagopal, preprint
(1999) [cond-math/9904100].

\bibitem{Fleming}  Fleming W, 1977 Functions of Several Variables (Berlin:
Springer).

\bibitem{Abe}  S. Abe, ${\it Physica A}$ {\bf 269} (1999) 403.


\end{references}
\end{document}